\documentclass[twocolumn,english]{revtex4-1}
\usepackage[T1]{fontenc}
\usepackage[latin9]{inputenc}
\usepackage{color}
\usepackage{babel}
\usepackage{mathtools}
\usepackage{amsmath}
\usepackage{amssymb}
\usepackage{graphicx}
\usepackage[unicode=true,pdfusetitle,
 bookmarks=false,
 breaklinks=false,pdfborder={0 0 1},backref=false,colorlinks=true]
 {hyperref}
\hypersetup{
 pdfborderstyle=}

\makeatletter

\providecommand{\tabularnewline}{\\}

\usepackage{babel}
\usepackage{babel}

\makeatother

\begin{document}
\title{Topological Classification of Excitations in Quadratic Bosonic Systems}
\author{Zixian Zhou}
\affiliation{Shenzhen Institute for Quantum Science and Engineering and Department
of Physics, Southern University of Science and Technology, Shenzhen
518055, China}
\author{Liang-Liang Wan}
\affiliation{Shenzhen Institute for Quantum Science and Engineering and Department
of Physics, Southern University of Science and Technology, Shenzhen
518055, China}
\author{Zhi-Fang Xu}
\email{xuzf@sustech.edu.cn}

\affiliation{Shenzhen Institute for Quantum Science and Engineering and Department
of Physics, Southern University of Science and Technology, Shenzhen
518055, China}
\begin{abstract}
We investigate the topological classification of excitations in quadratic
bosonic systems with an excitation band gap. Time-reversal, charge-conjugation,
and parity symmetries in bosonic systems are introduced to realize
a ten-fold symmetry classification. We find a specific decomposition
of the quadratic bosonic Hamiltonian and use it to prove that each
quadratic bosonic system is homotopic to a direct sum of two single-particle
subsystems. The topological classification table is thus derived via
inheriting from that of Atland-Zirnbauer classes and unique topological
phases of bosons are predicted. Finally, concrete topological models
are proposed to demonstrate the peculiarity of bosonic excitations. 
\end{abstract}
\maketitle

\section{Introduction}

Searching for topological phases of a many-body system with specific
symmetries becomes an important issue in both condensed matter and
cold atom physics. As a milestone work, gapped free-fermion systems
including the Chern insulator \cite{Haldane1988PhysRevLett}, topological
insulator \cite{Kane2005PhysRevLett,Kane2005PhysRevLett2,Bernevig2006Science,Fu2007PhysRevB}
and topological superconductor \cite{Read2000PhysRevB,Kitaev2001PU,Mourik2012Science,Sato2017RPP}
are categorized into ten symmetry classes according to the time-reversal,
charge-conjugation, and chiral symmetries \cite{Altland1997PhysRevB,Zirnbauer2010arXiv},
which is known as the Atland-Zirnbauer (AZ) classification. The relevant
topological phases are classified by a periodic table in the framework
of K-theory \cite{Stone2011JPA,Kitaev2009AIP,Zhao2013PhysRevLett,Chiu2016RevModPhys}.
Recently, the concept of topological phase has also been extended
to dynamical \cite{Chang2018PhysRevB,Yang2018PhysRevB,Gong2018PRL,Qiu2018PhysRevA}
and open quantum-mechanical systems \cite{Shen2018PhysRevLett,Kawabata2018PhysRevB,Yao2018PhysRevLett0,Yao2018PhysRevLett,Kawabata2019NatCommun}.
The classifications of Floquet topological insulator \cite{Roy2017PRB}
and non-Hermitian systems \cite{Gong2018PhysRevX,Kawabata2019PRX}
are also established based on similar classification principles.

Parallel to the fermionic insulator, topological phases also emerge
from excitations of a bosonic system. They not only are attainable
from the simulation of single-particle topological bands \cite{Haldane2008PhysRevLett,Wang2008PhysRevLett,Chong2008PhysRevB,Hafezi2011NatPhys,Hafezi2013NatPhotonics,Poo2011PhysRevLett,Poo2016SR,Skirlo2014PhysRevLett,Skirlo2015PhysRevLett,Wang2009Nature,Aidelsburger2013PhysRevLett,Aidelsburger2014NatPhys}
but also are discovered in peculiar bosonic systems without fermionic
analogy. It is reported that the bosonic Bogoliubov-de Gennes (BdG)
model containing two-boson annihilation/creation interactions \cite{Ring1980,Rossignoli2005PhysRevA,Blaizot1986}
is capable of hosting excitation band with non-vanishing Chern number
or $\mathbb{Z}_{2}$ index, which is realizable in magnonic crystals
\cite{Shindou2013PRB,Chisnell2015PRL,RoldanMolina2016NJP,Diaz2019PRL,Kondo2019PysRevB,Kondo2019arXiv},
nonlinear bosonic systems \cite{Bardyn2016}, photonic crystals \cite{Peano2016NatCommun},
and ultracold bosonic atoms in optical lattices \cite{Engelhardt2015PhysRevA,Xu2016PhysRevLett,DiLiberto2016,Luo2018}.
These bosonic excitation modes are obtained by a pseudo-unitary diagonalization,
i.e., Bogoliubov transformation, which keeps the bosonic commutation
relation \cite{Colpa1986PA,Simon1999JMP}. Thus, the topological obstruction,
i.e., the energy gap, is defined in an exotic way. Moreover, the stability
of bosonic Hamiltonians requires the semi-positive definiteness that
brings a limitation on the symmetries. These facts suggest that the
symmetry and topology of bosonic systems are qualitatively different
from those in the fermionic cases. Therefore, the bosonic excitations
are expected to generate peculiar topological phases beyond the common
AZ classification.

Then a problem naturally arises: It remains unclear if the topological
phases of bosonic excitation exist in other dimensions and symmetry
classes. Hence it is of great necessity to achieve their symmetry
and topological classifications.

In this paper, we focus on excitations of a quadratic bosonic system
(QBS) with an excitation band gap and systematically classify their
topological phases according to symmetries. We firstly inherit the
AZ classification scheme to introduce the time reversal, charge conjugation,
and their composite interpreted as parity and generate a ten-fold
symmetry classification for the QBS. We then explore the topological
structure of the bosonic Hamiltonian via a specific decomposition
which reveals that each QBS is homotopic to a direct sum of two single-particle
subsystems. Therefore, the classification table is derived via the
periodic table of AZ classes. We further apply these results to predict
unique topological phases of bosons and construct concrete bosonic
models without fermionic single-particle counterpart. Our work opens
a route to explore the topological phases and effects of bosons.

The paper is organized as follows: In Sec. \ref{sec:QBS}, we put
forward the model of QBS and give the symmetry classification. The
decomposition of bosonic Hamiltonian is derived. In Sec. \ref{sec:top},
the topological classification of excitations in QBS is made and the
topological invariants are discussed. In Sec. \ref{sec:Exam}, concrete
examples of interaction-driven topological phases are constructed.
In Sec. \ref{sec:conc}, conclusions are made.

\section{Quadratic bosonic system\label{sec:QBS}}

\subsection{Model}

We consider a quadratic Hamiltonian $\mathcal{H}=\sum_{k}\phi^{\dagger}\left(k\right)H\left(k\right)\phi\left(k\right)$
composed by bosonic field operators $\phi^{\dagger}\left(k\right)=\left(\begin{array}{cc}
\boldsymbol{a}^{\dagger}\left(k\right) & \boldsymbol{b}\left(-k\right)\end{array}\right)$ and an Hermitian matrix $H\left(k\right)$ which is continuous with
regard to wave vector $k$. Here, $\boldsymbol{a}^{\dagger}=\left(\begin{array}{ccc}
a_{1}^{\dagger} & \cdots & a_{N}^{\dagger}\end{array}\right)$ and $\boldsymbol{b}=\left(\begin{array}{ccc}
b_{1} & \cdots & b_{N^{\prime}}\end{array}\right)$ are bosonic creation and annihilation operators, respectively. The
field operators obey bosonic commutation relations 
\begin{equation}
\left[\phi_{i}\left(k\right),\phi_{j}^{\dagger}\left(k\right)\right]=\tau_{ij},~\tau=\left(\begin{array}{cc}
\mathbb{I}_{N}\\
 & -\mathbb{I}_{N^{\prime}}
\end{array}\right),\label{eq:comm}
\end{equation}
where $\mathbb{I}_{N}$ denotes $N\times N$ identity matrix. To stabilize
the system, $H\left(k\right)$ is required to be semi-positive definite.
This general QBS has included the single-particle system ($\boldsymbol{b}\left(k\right)$
vanishing) \cite{Wang2008PhysRevLett,Skirlo2014PhysRevLett,Skirlo2015PhysRevLett,Aidelsburger2013PhysRevLett,Aidelsburger2014NatPhys,Yan2018PhysRevLett}
and the widely studied bosonic BdG system ($b_{i}\left(k\right)=a_{i}\left(k\right)$,
$N=N^{\prime}$) \cite{Shindou2013PRB,Bardyn2016,Peano2016NatCommun,Engelhardt2015PhysRevA,Xu2016PhysRevLett,DiLiberto2016,Luo2018}.

We aim to investigate the excitation bands on top of a bosonic ground
state, which are solved via a linear transformation $\phi\left(k\right)=V\left(k\right)\psi\left(k\right)$.
To satisfy the bosonic commutation relation Eq. (\ref{eq:comm}),
the transformation matrix needs to obey $V^{\dagger}\tau V=\tau$
and forms a pseudo-unitary group $U\left(N,N^{\prime}\right)$. As
a mathematical theorem \cite{Simon1999JMP}, each positive definite
Hermitian matrix $H$ is pseudo-unitarily congruent to a positive
diagonal matrix $\Lambda$, i.e., 
\begin{equation}
V^{\dagger}HV=\Lambda,~V\in U\left(N,N^{\prime}\right).\label{eq:diag}
\end{equation}
Then the positive definite Hamiltonian takes a decoupled form $\mathcal{H}=\sum_{k}\psi^{\dagger}\left(k\right)\Lambda\left(k\right)\psi\left(k\right)$
with excitation modes $\psi\left(k\right)$ and energy spectra $\Lambda\left(k\right)$.
The Hamiltonian with zero excitation modes can be regarded as a limit
case and suits the same treatment. To generate a topological obstruction,
we assume an excitation band gap such that several energy bands are
always lower than the others. It leads to the appearance of bosonic
topological bands which cannot be continuously mapped to a trivial
flat-band model when the gap keeps open and the symmetries keep invariant.

It is worthy to mention that our subject is completely different from
the classification of bosonic topological orders \cite{Lan2018PRX,Lan2019PRX}.
The QBS in our consideration concerns the excitation band structure
in momentum space which will be classified by K-theory. In contrast,
the topological orders reflect the (real space) long-range entanglement
of ground states which are classified by tensor catogory theory.

\subsection{Symmetry classification}

For the symmetry classification of the QBS, we inherit the AZ classification
scheme to introduce time-reversal $\mathcal{T}$, charge-conjugation
$\mathcal{C}$, and their composite $\mathcal{P}=\mathcal{T}\cdot\mathcal{C}$
symmetries. Time-reversal operator is anti-unitary ($\mathcal{T}i\mathcal{T}^{-1}=-i$)
and defined by 
\begin{equation}
\mathcal{T}\phi_{i}\left(k\right)\mathcal{T}^{-1}:=\left(U_{T}\right)_{ij}\phi_{j}\left(-k\right),
\end{equation}
where $U_{T}$ is a unitary matrix. The charge-conjugation operator
is only adaptive to the case of $N=N^{\prime}$ and capable of reversing
the sign of the charge $\mathcal{Q}=\sum_{i,k}\left(a_{i}^{\dagger}a_{i}-b_{i}^{\dagger}b_{i}\right)$,
i.e., $\mathcal{C}\mathcal{Q}\mathcal{C}^{-1}=-\mathcal{Q}$. It is
unitary and defined by 
\begin{equation}
\mathcal{C}\phi_{i}\left(k\right)\mathcal{C}^{-1}:=\left(U_{C}^{\ast}\right)_{ij}\phi_{j}^{\dagger}\left(-k\right),
\end{equation}
where $U_{C}$ is also a unitary matrix. The $\mathcal{P}$ operator
as a composite operator is anti-unitary and satisfies 
\begin{equation}
\mathcal{P}\phi_{i}\left(k\right)\mathcal{P}^{-1}:=\left(U_{P}^{\ast}\right)_{ij}\phi_{j}^{\dagger}\left(k\right),
\end{equation}
where $U_{P}=U_{C}^{\ast}U_{T}$. According to these definitions,
an $\mathcal{O}$-symmetric Hamiltonian ($\mathcal{O}\mathcal{H}\mathcal{O}^{-1}=\mathcal{H}$)
where $\mathcal{O}=\mathcal{T},~\mathcal{C},~\mathcal{P}$ satisfies
the following constraint, 
\begin{equation}
TH\left(k\right)T^{-1}=H\left(-k\right),~T=U_{T}^{-1}\mathcal{K},\label{eq:T}
\end{equation}
\begin{equation}
CH\left(k\right)C^{-1}=H\left(-k\right),~C=U_{C}^{-1}\mathcal{K},\label{eq:C}
\end{equation}
\begin{equation}
PH\left(k\right)P^{-1}=H\left(k\right),~P=U_{P}^{-1}.\label{eq:P}
\end{equation}
Here, $\mathcal{K}$ denotes the complex conjugation. The redefined
symmetry operators $T,C,P$ act on $H\left(k\right)$ instead of $\mathcal{H}$.
Notably, $T,C$ are anti-unitary and $P$ becomes unitary.

Actually, the bosonic commutation relation assigns intrinsic structures
to the symmetry operators. From the $\mathcal{O}$-action upon Eq.
(\ref{eq:comm}), i.e., $\left[\mathcal{O}\phi_{i}\mathcal{O}^{-1},\mathcal{O}\phi_{j}^{\dagger}\mathcal{O}^{-1}\right]=\tau_{ij}$,
we find 
\begin{equation}
T\tau T^{-1}=\tau,~C\tau C^{-1}=-\tau,~P\tau P^{-1}=-\tau.\label{eq:struct}
\end{equation}
Furthermore, we assume that $\mathcal{T}$, $\mathcal{C}$, $\mathcal{P}$
are involutive operators (twice action making any system return to
itself) and achieve the following properties by Schur's lemma, 
\begin{equation}
T^{2}=\pm\mathbb{I},~C^{2}=\pm\mathbb{I},~P^{2}=e^{i\varphi}\mathbb{I}.\label{eq:inv}
\end{equation}
The real phase $\varphi$ can be dropped by a global $U\left(1\right)$
gauge transformation in advance. Consequently, the representation
matrices of $U_{T,C,P}$ satisfying Eqs. (\ref{eq:struct}--\ref{eq:inv})
take the following forms, 
\begin{equation}
U_{T}=\left(\begin{array}{cc}
u_{t}\\
 & u_{t}^{\prime}
\end{array}\right),~U_{C}=\left(\begin{array}{cc}
 & \pm u_{c}\\
u_{c}^{T}
\end{array}\right),~U_{P}=\left(\begin{array}{cc}
 & u_{p}\\
u_{p}^{\dagger}
\end{array}\right),\label{eq:U}
\end{equation}
where $u_{t}$, $u_{t}^{\prime}$, $u_{c}$, and $u_{p}$ are all
unitary matrices.

Comparing to the AZ classification \cite{Altland1997PhysRevB,Chiu2016RevModPhys},
we find three different features: (1) the symmetric constraints for
$T$ and $C$ given by Eqs. (\ref{eq:T}--\ref{eq:C}) take the same
form, (2) $T$ and $C$ are identified by their relations to $\tau$
according to Eq. (\ref{eq:struct}), and (3) $P$ should be named
parity according to Eq. (\ref{eq:P}), in contrast to the fermionic
case where $T\cdot C$ is interpreted as chirality. This is because
the fermionic anti-commutation relation is replaced by the bosonic
commutation relation.

Based on the presence or absence of these three symmetries, the symmetry
classification of the QBS is listed in the first four columns of Tab.
\ref{tab:topology}. This result resembles the AZ ten-fold symmetry
classification for fermions. Nevertheless, the symmetry classes correspond
to repeated classifying spaces as revealed later. Thus, different
labels compared to Cartan's are used.

\begin{table*}
\caption{\label{tab:topology}Symmetry and topological classification of QBS,
where the $k$-space is sphere $S^{d}$. In the first four columns,
``C, R, H'' marks the complex ($\mathbb{C}$), real ($\mathbb{R}$),
quaternionic ($\mathbb{H}$) classes whose $T$ symmetry reads $0,+,-$,
respectively; ``I, II'' marks the $C$ symmetry being $+,-$, respectively,
and ``III'' marks the class with $P$ symmetry only. Here 0 refers
to the absence of symmetry, 1 refers to the presence of $P$ symmetry,
and $\pm$ refers to the presence of $T$ or $C$ symmetry with $T^{2}=\pm\mathbb{I}$
or $C^{2}=\pm\mathbb{I}$. The classifying space for the complex,
real, quaternionic classes is given by $\mathcal{C}_{0}=\mathbb{Z}\times BU$,
$\mathcal{R}_{0}=\mathbb{Z}\times BO$, $\mathcal{R}_{4}=\mathbb{Z}\times BSp$
with Bott periodicity 2, 8, 8, respectively. In the last eight columns,
entries like $0$, $\mathbb{Z}$ and $\mathbb{Z}_{2}$ represent the
possible topological phases.}
\begin{tabular}{ccccccccccccc}
\hline 
Label & $T$ & $C$ & $P$ & Classifying space & $d=0$ & $d=1$ & $d=2$ & $d=3$ & $d=4$ & $d=5$ & $d=6$ & $d=7$\tabularnewline
\hline 
C & 0 & 0 & 0 & $\mathcal{C}_{0}\times\mathcal{C}_{0}$ & $\mathbb{Z}\oplus\mathbb{Z}$ & $0$ & $\mathbb{Z}\oplus\mathbb{Z}$ & $0$ & $\mathbb{Z}\oplus\mathbb{Z}$ & $0$ & $\mathbb{Z}\oplus\mathbb{Z}$ & $0$\tabularnewline
CI & 0 & + & 0 & $\mathcal{C}_{0}$ & $\mathbb{Z}$ & $0$ & $\mathbb{Z}$ & $0$ & $\mathbb{Z}$ & $0$ & $\mathbb{Z}$ & $0$\tabularnewline
CII & 0 & - & 0 & $\mathcal{C}_{0}$ & $\mathbb{Z}$ & $0$ & $\mathbb{Z}$ & $0$ & $\mathbb{Z}$ & $0$ & $\mathbb{Z}$ & $0$\tabularnewline
CIII & 0 & 0 & 1 & $\mathcal{C}_{0}$ & $\mathbb{Z}$ & $0$ & $\mathbb{Z}$ & $0$ & $\mathbb{Z}$ & $0$ & $\mathbb{Z}$ & $0$\tabularnewline
R & + & 0 & 0 & $\mathcal{R}_{0}\times\mathcal{R}_{0}$ & $\mathbb{Z}\oplus\mathbb{Z}$ & $0$ & $0$ & $0$ & $\mathbb{Z}\oplus\mathbb{Z}$ & $0$ & $\mathbb{Z}_{2}\oplus\mathbb{Z}_{2}$ & $\mathbb{Z}_{2}\oplus\mathbb{Z}_{2}$\tabularnewline
RI & + & + & 1 & $\mathcal{R}_{0}$ & $\mathbb{Z}$ & $0$ & $0$ & $0$ & $\mathbb{Z}$ & $0$ & $\mathbb{Z}_{2}$ & $\mathbb{Z}_{2}$\tabularnewline
RII & + & - & 1 & $\mathcal{R}_{0}$ & $\mathbb{Z}$ & $0$ & $0$ & $0$ & $\mathbb{Z}$ & $0$ & $\mathbb{Z}_{2}$ & $\mathbb{Z}_{2}$\tabularnewline
H & - & 0 & 0 & $\mathcal{R}_{4}\times\mathcal{R}_{4}$ & $\mathbb{Z}\oplus\mathbb{Z}$ & $0$ & $\mathbb{Z}_{2}\oplus\mathbb{Z}_{2}$ & $\mathbb{Z}_{2}\oplus\mathbb{Z}_{2}$ & $\mathbb{Z}\oplus\mathbb{Z}$ & $0$ & $0$ & $0$\tabularnewline
HI & - & + & 1 & $\mathcal{R}_{4}$ & $\mathbb{Z}$ & $0$ & $\mathbb{Z}_{2}$ & $\mathbb{Z}_{2}$ & $\mathbb{Z}$ & $0$ & $0$ & $0$\tabularnewline
HII & - & - & 1 & $\mathcal{R}_{4}$ & $\mathbb{Z}$ & $0$ & $\mathbb{Z}_{2}$ & $\mathbb{Z}_{2}$ & $\mathbb{Z}$ & $0$ & $0$ & $0$\tabularnewline
\hline 
\end{tabular}
\end{table*}

\subsection{Decomposition of Hamiltonian}

As a preliminary of topological analysis, we introduce a specific
decomposition of bosonic Hamiltonian $H\left(k\right)$ based on the
peculiar pseudo-unitary diagonalization.

Firstly, we figure out the topology of pseudo-unitary group $U\left(N,N^{\prime}\right)$.
According to the definition $V^{\dagger}\tau V=\tau$, each $V=\left(\begin{array}{cc}
v_{a} & v_{x}\\
v_{y} & v_{b}
\end{array}\right)\in U\left(N,N^{\prime}\right)$ satisfies

\begin{equation}
v_{a}^{\dagger}v_{a}-v_{y}^{\dagger}v_{y}=\mathbb{I},~v_{b}^{\dagger}v_{b}-v_{x}^{\dagger}v_{x}=\mathbb{I},~v_{a}^{\dagger}v_{x}-v_{y}^{\dagger}v_{b}=0.
\end{equation}
As $\det v_{a}^{\dagger}v_{a}\geq1$ and $\det v_{b}^{\dagger}v_{b}\geq1$,
we infer that $v_{a}$ and $v_{b}$ are invertible matrices. This
enables us to denote $r=v_{x}v_{b}^{-1}=\left(v_{y}v_{a}^{-1}\right)^{\dagger}$
to reduce the above equations to 
\begin{equation}
\mathbb{I}-rr^{\dagger}=\left(v_{a}v_{a}^{\dagger}\right)^{-1},~\mathbb{I}-r^{\dagger}r=\left(v_{b}v_{b}^{\dagger}\right)^{-1}.
\end{equation}
Then the pseudo-unitary matrix can be recast as 
\begin{eqnarray}
V & = & \left(\begin{array}{cc}
\mathbb{I} & r\\
r^{\dagger} & \mathbb{I}
\end{array}\right)\left(\begin{array}{cc}
v_{a}\\
 & v_{b}
\end{array}\right)\\
 & = & \left(\begin{array}{cc}
\mathbb{I} & r\\
r^{\dagger} & \mathbb{I}
\end{array}\right)\left(\begin{array}{cc}
\left(\mathbb{I}-rr^{\dagger}\right)^{-\frac{1}{2}}u_{a}\\
 & \left(\mathbb{I}-r^{\dagger}r\right)^{-\frac{1}{2}}u_{b}
\end{array}\right),
\end{eqnarray}
where $u_{a,b}=\left(v_{a,b}v_{a,b}^{\dagger}\right)^{-\frac{1}{2}}v_{a,b}$
are unitary matrices. Defining 
\begin{equation}
W=\tanh^{-1}\left(\begin{array}{cc}
 & r\\
r^{\dagger}
\end{array}\right)=\left(\begin{array}{cc}
 & w\\
w^{\dagger}
\end{array}\right),
\end{equation}
we can simplify the above expression as 
\begin{equation}
V=e^{W}\left(\begin{array}{cc}
u_{a}\\
 & u_{b}
\end{array}\right).\label{eq:V}
\end{equation}
Therefore, a pseudo-unitary matrix $V$ is composed by an $N\times N^{\prime}$
matrix $w$ and two unitary matrices $u_{a,b}$, which implies that
$U\left(N,N^{\prime}\right)$ is homeomorphic to $\mathbb{C}^{NN^{\prime}}\times U\left(N\right)\times U\left(N^{\prime}\right)$.

We then introduce the decomposition of $H\left(k\right)$. By substituting
Eq. (\ref{eq:V}) to Eq.(\ref{eq:diag}), we can recast $H\left(k\right)$
as 
\begin{equation}
H\left(k\right)=e^{-W}H_{0}e^{-W},\label{eq:decom}
\end{equation}
where 
\begin{equation}
H_{0}\left(k\right)=\left(\begin{array}{cc}
u_{a}\\
 & u_{b}
\end{array}\right)\Lambda\left(\begin{array}{cc}
u_{a}^{\dagger}\\
 & u_{b}^{\dagger}
\end{array}\right)=\left(\begin{array}{cc}
h_{0}\\
 & h_{0}^{\prime}
\end{array}\right).
\end{equation}
This implies that a QBS is decomposed into two effective single-particle
subsystems $h_{0}\left(k\right),~h_{0}^{\prime}\left(k\right)$ with
a coupling generator $w\left(k\right)$, and its excitation spectra
are identical to those of the subsystems. Conversely, $W\left(k\right)$
and $H_{0}\left(k\right)$ can also be expressed in term of $H\left(k\right)$.
We notice the identity 
\begin{equation}
\tau H\tau=e^{W}H_{0}e^{W}=e^{2W}He^{2W}
\end{equation}
and find the unique solution 
\begin{eqnarray}
e^{2W} & = & H^{-\frac{1}{2}}\left(H^{\frac{1}{2}}\tau H\tau H^{\frac{1}{2}}\right)^{\frac{1}{2}}H^{-\frac{1}{2}},\label{eq:W}\\
H_{0} & = & e^{W}He^{W}.\label{eq:sing}
\end{eqnarray}
Based on these formula, we know that $W\left(k\right)$ and $H_{0}\left(k\right)$
are continuous with respect to $k$ and obey the same symmetric constraints
as the original Hamiltonian $H\left(k\right)$.

\section{Topological classification\label{sec:top}}

\subsection{Homotopic property}

The topological feature of bosonic excitations is fully encoded in
the homotopic property of the elementary-excitation Hamiltonian. We
define the homotopy as follows: If $H\left(k\right)$ can be continuously
mapped to $H^{\prime}\left(k\right)$ without breaking the symmetry
and closing the excitation gap, we say that $H\left(k\right)$ and
$H^{\prime}\left(k\right)$ are homotopic, denoted as $H\left(k\right)\approx H^{\prime}\left(k\right)$.
Based on Eq. (\ref{eq:decom}), we construct a continuous series of
Hamiltonian, 
\begin{equation}
H_{\epsilon}\left(k\right)=e^{-\epsilon W\left(k\right)}H_{0}\left(k\right)e^{-\epsilon W\left(k\right)},~\epsilon\in\left[0,1\right],
\end{equation}
which evidently shares the same symmetries and energy spectra of $H\left(k\right)$.
Therefore, we immediately achieve a homotopic relation 
\begin{equation}
H\left(k\right)=H_{1}\left(k\right)\approx H_{0}\left(k\right)=h_{0}\left(k\right)\oplus h_{0}^{\prime}\left(k\right).
\end{equation}
This means that a QBS is topologically equivalent to two gapped fermion-like
subsystems which belong to AZ classes. As an intuitive comprehension,
each $V=e^{W}\left(u_{a}\oplus u_{b}\right)$ can be continuously
mapped to $u_{a}\oplus u_{b}$ through linearly decreasing $W$ to
zero. This mapping forms a deformation retraction of $U\left(N,N^{\prime}\right)$
onto $U\left(N\right)\times U\left(N^{\prime}\right)$ \cite{DefRe}
which implies that the two Lie groups are homotopy equivalent. The
Hamiltonian $H\left(k\right)$ diagonalized by $V\left(k\right)$
can be retracted to $H_{0}\left(k\right)$ diagonalized by $u_{a}\oplus u_{b}$,
where the spectra and topological features remain unchanged during
the retraction.

Next, we need to figure out the intrinsic structure and interrelation
between two subsystems. When there is only $T$ symmetry or no symmetry,
two subsystems are fully independent. Thus, the topological classification
of $H\left(k\right)$ is given by that of $h_{0}\left(k\right)\oplus h_{0}^{\prime}\left(k\right)$.
When $H\left(k\right)$ is $C$-invariant or $P$-invariant, there
are constraints among two subsystems as given by Eqs. (\ref{eq:C}--\ref{eq:P}),
i.e., 
\begin{equation}
u_{c}^{-1}h_{0}^{\ast}\left(-k\right)u_{c}=h_{0}^{\prime}\left(k\right),~u_{p}^{-1}h_{0}\left(k\right)u_{p}=h_{0}^{\prime}\left(k\right),
\end{equation}
where Eq. (\ref{eq:U}) has been used. This implies that the topological
classification of $H\left(k\right)$ is fully determined by $h_{0}\left(k\right)$.
In other word, the $C$ and $P$ symmetries in the QBS establish the
relation between $h_{0}\left(k\right)$ and $h_{0}^{\prime}\left(k\right)$
rather than confine their intrinsic structures. Therefore, subsystems
$h_{0}\left(k\right)$ and $h_{0}^{\prime}\left(k\right)$ only contain
the same $T$ symmetry as the original system $H\left(k\right)$,
attributed to class A ($T=0$), class AI ($T^{2}=+\mathbb{I}$), or
class AII ($T^{2}=-\mathbb{I}$) in the AZ classification \cite{Altland1997PhysRevB}.

Now we are able to finish the classification of excitations in QBS
by directly extending the classification of AZ classes, which is summarized
in Tab. \ref{tab:topology}. The details related to the profound K-theory
are presented in \ref{sec:Class_AZ}. In this table, we see that the
bosonic BdG systems labeled by class CI has the same classification
of Chern insulator, i.e., class A of AZ scheme. It coincides with
the results given by Refs. \cite{Peano2018JMP,Lein2019PRB} and provides
a verification.

Besides, it is worthy to clarify the relation between stable bosonic
Hamiltoniain $H\left(k\right)$ and pseudo-Hermitian Hamiltonian \cite{Kawabata2019PRX}.
Although each $H\left(k\right)$ is mapped to a pseudo-Hermitian Hamiltonian
$\tau H\left(k\right)$ which has real spectrum via similarity diagonalization,
not each general pseudo-Hermitian Hamiltonian with complex spectrum
corresponds to a bosonic one. Therefore, the classification of $H\left(k\right)$
cannot be replaced by that of pseudo-Hermitian Hamiltonian.

\subsection{Topological invariants}

Although the topological phasesare classified, we still need characteristic
numbers to distinguish different topological phases in each symmetry
class. Since the topology of $H\left(k\right)$ is fully determined
by $h_{0}\left(k\right)$ and $h_{0}^{\prime}\left(k\right)$, the
topological invariants of $H\left(k\right)$ are just given by those
of the fermion-like subsystems $h_{0}\left(k\right)$ and $h_{0}^{\prime}\left(k\right)$.
Inherited from the fermionic case, the $\mathbb{Z}$-type invariants
in Tab. \ref{tab:topology} are essentially the Chern numbers of the
bands below the gap (for even $d$), and the $\mathbb{Z}_{2}$-type
invariants are interpreted as the Chern-Simons invariants for odd
$d$ or the Fu-Kane invariants for even $d$ \cite{Chiu2016RevModPhys}.
We completely list the characteristic numbers in Tab. \ref{tab:character}
and provide their formulas in \ref{sec:numbers}.

\begin{table}
\caption{\label{tab:character}Characteristic numbers as interpretations of
the topological invariants in Tab. \ref{tab:topology}. Here ``CN,
FK, CS'' refers to the Chern number, Fu-Kane invariant, and Chern-Simons
invariant, respectively.}
\begin{tabular}{ccc}
\hline 
Invariant & even $d$ & odd $d$\tabularnewline
\hline 
$\mathbb{Z}$ & CN for $h_{0}$ & /\tabularnewline
$\mathbb{Z}\oplus\mathbb{Z}$ & CN for $h_{0}$ and $h_{0}^{\prime}$ & /\tabularnewline
$\mathbb{Z}_{2}$ & FK for $h_{0}$ & CS for $h_{0}$\tabularnewline
$\mathbb{Z}_{2}\oplus\mathbb{Z}_{2}$ & FK for $h_{0}$ and $h_{0}^{\prime}$ & CS for $h_{0}$ and $h_{0}^{\prime}$\tabularnewline
\hline 
\end{tabular}
\end{table}

We further predict unique topological phases of bosonic excitations
from Tab. \ref{tab:topology}, including (1) asymmetric system in
class C for $d=2,4$ and (2) $T$-invariant system in class H for
$d=2,3$. Their topological structures are characterized by a pair
of Chern numbers and a pair of $\mathbb{Z}_{2}$ indices, respectively,
which double the results of their fermionic counterparts. The relevant
models need to contain independent bosonic operators $\boldsymbol{a}$
and $\boldsymbol{b}$ which exceed the common BdG system. The simplest
model consists of two single-particle subsystems 
\begin{equation}
\mathcal{H}_{a}=\sum_{k}\boldsymbol{a}^{\dagger}\left(k\right)h_{a}\left(k\right)\boldsymbol{a}\left(k\right),~\mathcal{H}_{b}=\sum_{k}\boldsymbol{b}\left(k\right)h_{b}\left(k\right)\boldsymbol{b}^{\dagger}\left(k\right)
\end{equation}
and the (on-site) two-mode squeezing coupling \cite{Scully1997} 
\begin{equation}
\mathcal{H}_{x}=\sum_{k}\boldsymbol{a}^{\dagger}\left(k\right)h_{x}\boldsymbol{b}^{\dagger}\left(-k\right)+\mathrm{h.c.},
\end{equation}
i.e., $\mathcal{H}=\mathcal{H}_{a}+\mathcal{H}_{b}+\mathcal{H}_{x}$.
When $d=1$, all the symmetry classes are trivial because the disappearance
of chirality invalidates the winding number that labels the topological
phase. Besides, topological phases may arise in every symmetry class
for $d=4$ as predicted in Tab. \ref{tab:topology}, which are possibly
implemented in artificial dimensions.

\section{Interaction-driven topological phases\label{sec:Exam}}

As an application of our discovery, the technique of Hamiltonian decomposition
allows us to construct peculiar bosonic models without fermionic counterpart.
The typical feature of a QBS is the two-boson annihilation/creation
interactions $ba$/$a^{\dagger}b^{\dagger}$ which may arise from
the two-photon squeezing in a photonic crystal or the atomic interaction
of ultracold atoms. It is possible to impose two-boson interactions
on the trivial single-particle parts to make the complete Hamiltonian
topological, generating interaction-driven topological phases.

\subsection{$\mathbb{Z}$-type model}

The first example that we consider is a 2D bosonic BdG Hamiltonian
\begin{equation}
H_{\mathrm{BdG}}\left(k\right)=\left(\begin{array}{cc}
h_{a}\left(k\right) & h_{x}\left(k\right)\\
h_{x}^{\ast}\left(-k\right) & h_{a}^{\ast}\left(-k\right)
\end{array}\right).
\end{equation}
This model possesses $C$ symmetry with $C^{2}=+\mathbb{I}$ and is
attributed to class CI. According to Tabs. \ref{tab:topology} and
\ref{tab:character}, its topological phases are classified by $\mathbb{Z}$
and characterized by a Chern number in even dimensions. This result
has been already achieved by Ref. \cite{Peano2018JMP}. Here, we propose
the trivial block matrices as follows, 
\begin{equation}
h_{a}\left(k\right)=\sigma^{1}\sin k_{1}+\sigma^{2}\sin k_{2}-\mu\mathbb{I},~(\mu<0),
\end{equation}
\begin{equation}
h_{x}\left(k\right)=\sigma^{3}\left(m+\cos k_{1}+\cos k_{2}\right)+\xi\mathbb{I},~(m,\xi\in\mathbb{C}),
\end{equation}
where $\sigma^{1,2,3}$ are Pauli matrices and $k_{1,2}\in\left[0,2\pi\right]$
form a 2D torus.

To achieve the topological phase diagram, we consider another homotopy
equivalent Hamiltonian which is given by 
\begin{equation}
H\left(k;\Omega\right)=H_{\mathrm{BdG}}\left(k\right)+\Omega\mathbb{I}\approx H_{\mathrm{BdG}}\left(k\right).
\end{equation}
Its analytical result can be solved in limit $\Omega\rightarrow+\infty$.
Since $H\left(k;\Omega\right)$ obviously keeps the symmetries of
$H_{\mathrm{BdG}}\left(k\right)$, we just need to prove that the
gap keeps open while $\Omega$ changes from $0$ to $+\infty$. (The
spectra of $H_{\mathrm{BdG}}\left(k\right)+\Omega\mathbb{I}$ are
not simply $\Lambda\left(k\right)+\Omega\mathbb{I}$ due to the pseudo-unitary
diagonalization.) The proof starts from the inversion symmetry and
the $C$ symmetry, i.e., 
\begin{equation}
U_{I}^{-1}H\left(-k;\Omega\right)U_{I}=H\left(k;\Omega\right),~U_{I}=\mathbb{I}_{2}\otimes\sigma^{3},
\end{equation}
\begin{equation}
U_{C}^{-1}H^{\ast}\left(-k;\Omega\right)U_{C}=H\left(k;\Omega\right),~U_{C}=\sigma^{1}\otimes\mathbb{I}_{2}.
\end{equation}
These symmetric structures result in the two-fold degeneracy of the
spectra. When the gap is closed at a certain $k$, energy bands intersect
at one point, i.e., $H_{0}\left(k;\Omega\right)=\lambda\mathbb{I}$.
According to the Hamiltonian decomposition Eq. (\ref{eq:decom}),
it is equivalent to 
\begin{equation}
\left(\tau H\right)^{2}=\lambda^{2}\mathbb{I}.
\end{equation}
After substituting the expression of $H\left(k;\Omega\right)$ into
it, we reduce the gap-closing condition to 
\begin{equation}
\sin k_{j}=0,~\xi^{\ast}\left(m+\cos k_{1}+\cos k_{2}\right)+\mathrm{c.c.}=0
\end{equation}
which are independent of $\Omega$. As a result, $H\left(k;\Omega\right)$
is gapped as long as $H_{\mathrm{BdG}}\left(k\right)$ is gapped.
Thus, the homotopic relation $H\left(k;\Omega\right)\approx H_{\mathrm{BdG}}\left(k\right)$
is proved. Next, we are able to apply perturbation theory in limit
of $\Omega\rightarrow+\infty$. After tedious calculation, Eqs. (\ref{eq:W}--\ref{eq:sing})
are reduced to 
\begin{equation}
h_{0}\left(k;\Omega\right)=h_{a}\left(k\right)-\frac{1}{2\Omega}h_{x}h_{x}^{\dagger}\left(k\right)+\Omega\mathbb{I}.
\end{equation}
We can infer that it is a Chern insulator \cite{Qi2006PhysRevB} whose
Chern number is given by 
\begin{eqnarray}
\mathrm{CN}= & \mathrm{sgn}\left(\mathrm{Re}\xi^{\ast}m\right), & 0<\left|\mathrm{Re}\xi^{\ast}m\right|<2\left|\mathrm{Re}\xi\right|\\
= & 0. & 2\left|\mathrm{Re}\xi\right|<\left|\mathrm{Re}\xi^{\ast}m\right|\label{eq:CN}
\end{eqnarray}

\begin{figure}
\centering\includegraphics[width=0.6\columnwidth]{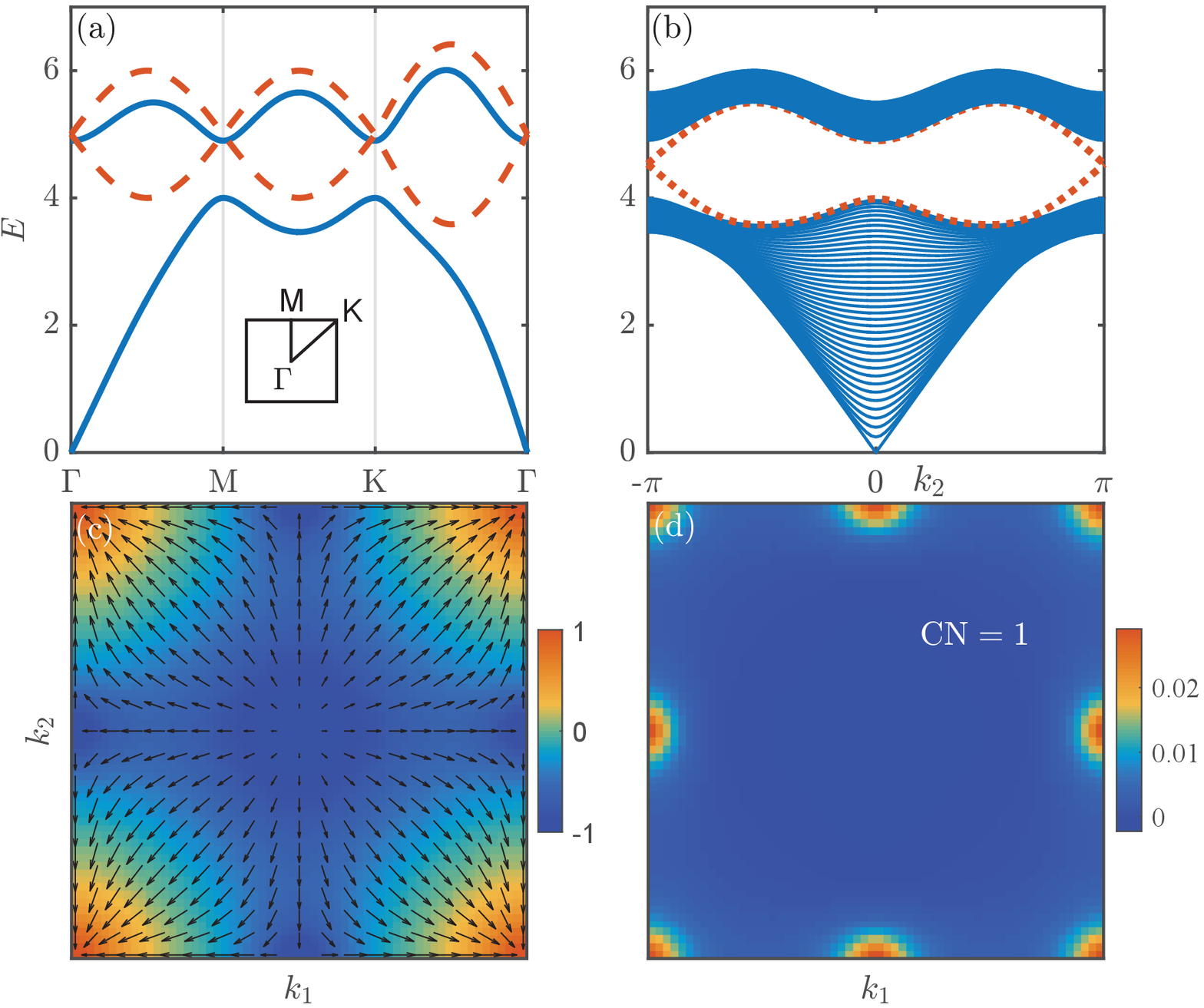}

\caption{(color online). \label{fig:model} (a) Excitation spectra $E\left(k\right)$
(blue solid lines) and energy spectra of $h_{a}\left(k\right)$ (red
dashed lines). (b) Excitation spectra $E\left(k_{2}\right)$ for the
lattice model of $H_{\mathrm{BdG}}\left(k\right)$ in open-boundary
condition. (c) Spin texture $\boldsymbol{l}\left(k\right)/\left|\boldsymbol{l}\left(k\right)\right|$
with the arrow representing $\left(l_{1},l_{2}\right)/\left|\boldsymbol{l}\right|$
and the color representing $l_{3}/\left|\boldsymbol{l}\right|$. (d)
Berry curvature. Parameters: $\mu=-5$, $m=1$ and $\xi=2$.}
\end{figure}

We further numerically solve the effective single-particle via Eqs.
(\ref{eq:W}--\ref{eq:sing}). Assuming 
\begin{equation}
h_{0}\left(k\right)=\lambda\mathbb{I}+\boldsymbol{l}\cdot\boldsymbol{\sigma},
\end{equation}
where $\lambda\left(k\right)>0$ and $\boldsymbol{l}\left(k\right)\in\mathbb{R}^{3}\backslash0$,
we obtain values of $\lambda$ and $\boldsymbol{l}$ numerically.
The spin texture $\boldsymbol{l}/\left|\boldsymbol{l}\right|$ and
excitation spectra $E\left(k\right)=\lambda\pm\left|\boldsymbol{l}\right|$
for parameters $\mu=-5$, $\xi=2$ and $m=1$ are presented in Fig.
\ref{fig:model} (a) and (c), respectively. We infer that $E\left(k\right)$
in Fig. \ref{fig:model} (a) opens a gap in the presence of the two-boson
interactions. And the skyrmion in Fig. \ref{fig:model} (c) reflects
the non-triviality of the interaction-driven phase. We also present
the Berry curvature in Fig. \ref{fig:model} (d) and integrate it
to gain the Chern number. The numerical result $\mathrm{CN}=1$ coincides
with the analytical result of Eq. (\ref{eq:CN}). Finally, we consider
the lattice model of $H_{\mathrm{BdG}}\left(k\right)$ and adopt open
and periodic boundary conditions along two othorgonal directions,
respectively. Two edge modes inside the excitation gap are observed
in the excitation spectra as shown in Fig. \ref{fig:model} (b). The
presence of edge modes is closely related to the Chern number of the
bulk Hamiltonian $H\left(k\right)$ \cite{Peano2018JMP}, which is
similar to the fermionic case.

\subsection{$\mathbb{Z}_{2}$-type model}

We also propose a $\mathbb{Z}_{2}$-type topological phase in 2D and
3D based on the similar construction. We choose the BdG Hamiltonian
with block matrices 
\begin{equation}
h_{a}\left(k\right)=\sum_{j=1}^{d}\gamma^{j}\sin k_{j}-\mu\mathbb{I},~(\mu<0),
\end{equation}
\begin{equation}
h_{x}\left(k\right)=\gamma^{0}\left(m+\sum_{j=1}^{d}\cos k_{j}\right)+\xi\mathbb{I},~(m,\xi\in\mathbb{R}).
\end{equation}
Here $\gamma^{0},\gamma^{j}$ are the Clifford generators and $k_{j}\in\left[0,2\pi\right]$
form a $d$-dimensional torus ($d=2,3$). This model possesses all
the $T,C,P$ symmetries and belongs to class HI. Its topological phases
are classified by $\mathbb{Z}_{2}$ for both 2D and 3D. Similar to
the above analysis, the $C$ symmetry and Kramer degeneracy from $T^{2}=-\mathbb{I}$
symmetry causes spectral degeneracy, so that the spectra consist of
two separate bands. Then the gap-closing condition is also given by
$H_{0}\left(k;\Omega\right)=\lambda\mathbb{I}$ which is equivalent
to $\left(\tau H\right)^{2}=\lambda^{2}\mathbb{I}$. We can also prove
that $H_{\mathrm{BdG}}\left(k\right)+\Omega\mathbb{I}\approx H_{\mathrm{BdG}}\left(k\right)$
and obtain a perturbative result 
\begin{equation}
h_{0}\left(k\right)\approx h_{a}\left(k\right)-\frac{1}{2\Omega}h_{x}h_{x}^{\dagger}\left(k\right)+\Omega\mathbb{I}
\end{equation}
in the limit of $\Omega\rightarrow\infty$. This single-particle Hamiltonian
serves as a $T$-invariant topological insulator \cite{Bernevig2013}.
When $\xi\neq0$, topological non-trivial phases appear if $d-2<\left|m\right|<d$
and trivial phases emerge if $\left|m\right|>d$ or $\left|m\right|<d-2$.

\section{Conclusions\label{sec:conc}}

We have accomplished the symmetry and topological classification of
the QBS with an excitation band gap. Three basic symmetries in QBS
are introduced and the ten-fold symmetry classification is realized.
A specific decomposition of the elementary-excitation Hamiltonian
is applied to reveal its algebraic and topological structures. Then
the classification table of excitations in QBS is derived based on
that of AZ classes. Unique topological phases of bosons are discussed
and concrete bosonic models without fermionic counterpart are constructed.
Our work provides a framework to explore richer topological physics
of bosons.

The possible studies in future include the implementation of the predicted
topological phases in realistic systems, the extension of the classification
table by considering lattice symmetries, and a systematical investigation
on the bulk-edge correspondence in the bosonic case.
\begin{acknowledgments}
This work is supported by National Key R\&D Program of China (under
Grant No. 2018YFA0307200), the Key-Area Research and Development Program
of GuangDong Province (Grant No. 2019B030330001) and National Natural
Science Foundation of China (under Grant No. 11574100 and U1801661).
\end{acknowledgments}

\appendix

\section{Classification principle\label{sec:Class_AZ}}

We briefly introduce the classification principle of AZ classes for
free-fermion systems and extend it to our quadratic bosonic systems.
For the fermionic AZ classes, usually one first introduces the trivial
Hamiltonian 
\begin{equation}
\sigma=\left(\begin{array}{cc}
\mathbb{I}_{r}\\
 & -\mathbb{I}_{r}
\end{array}\right),
\end{equation}
which corresponds to a flat-band system with equal conduction and
valence bands. Next, we apply K-theory to define the stable equivalence
of single-particle Hamiltonians $h\left(k\right)$ and $h^{\prime}\left(k\right)$
as follows, 
\begin{equation}
h\left(k\right)\oplus\sigma\approx h^{\prime}\left(k\right)\oplus\sigma^{\prime},
\end{equation}
where $\sigma$ and $\sigma^{\prime}$ are independent trivial Hamiltonians
with unlimited matrix sizes. The stable equivalence studies the homotopy
of Hamiltonians whose intrinsic space is enlarged to $\infty$-dimension
by attaching flat bands. It is a looser condition than the original
homotopy such that the classification becomes much easier.

People usually regard the stably equivalent class $\left[h\left(k\right)\right]$
as a topological phase and achieve the classification by counting
out all the $\left[h\left(k\right)\right]$ for given symmetries and
$k$-space dimension \cite{Chiu2016RevModPhys,Kitaev2009AIP}. The
classification table is derived from the Bott periodicity theorem
\cite{Stone2011JPA}. The results of class A, AI, and AII are presented
in Tab. \ref{tab:AZ}, where the $k$-space is chosen as sphere $S^{d}$.
If the $k$-space becomes the usual torus $T^{d}$ in band theory,
the final result is the present answer plus some weak topological
invariants \cite{Kitaev2009AIP}.

\begin{table}
\caption{\label{tab:AZ}Topological classification of class A, AI, and AII.
The $k$-space is chosen as sphere $S^{d}$. In the second column,
symbol 0 refers to absence of symmetry and $\pm$ refers to the presence
of symmetry with $T^{2}=\pm\mathbb{I}$.}
\begin{tabular}{ccccccccccc}
\hline 
AZ class & $T$ & Classifying space & $d=$0 & 1 & 2 & 3 & 4 & 5 & 6 & 7\tabularnewline
\hline 
A & 0 & $\mathcal{C}_{0}=\mathbb{Z}\times BU$ & $\mathbb{Z}$ & $0$ & $\mathbb{Z}$ & $0$ & $\mathbb{Z}$ & $0$ & $\mathbb{Z}$ & $0$\tabularnewline
AI & + & $\mathcal{R}_{0}=\mathbb{Z}\times BO$ & $\mathbb{Z}$ & $0$ & $0$ & $0$ & $\mathbb{Z}$ & $0$ & $\mathbb{Z}_{2}$ & $\mathbb{Z}_{2}$\tabularnewline
AII & - & $\mathcal{R}_{4}=\mathbb{Z}\times BSp$ & $\mathbb{Z}$ & $0$ & $\mathbb{Z}_{2}$ & $\mathbb{Z}_{2}$ & $\mathbb{Z}$ & $0$ & $0$ & $0$\tabularnewline
\hline 
\end{tabular}
\end{table}

For the quadratic bosonic system, we naturally extend the trivial
Hamiltonian $O$ as follows, 
\begin{equation}
O=\left(\begin{array}{cc}
o_{\tau=+1}\\
 & o_{\tau=-1}
\end{array}\right),~o=\left(\begin{array}{cc}
E_{\mathrm{up}}\mathbb{I}_{r}\\
 & E_{\mathrm{low}}\mathbb{I}_{r}
\end{array}\right),
\end{equation}
where the block matrix $o_{\tau=\pm1}$ acting on the $\tau=\pm1$
eigenspace corresponds to a flat-band subsystem with equal upper and
lower excitation bands (the excitation bands above and below the gap).
Hence, attaching it on $H\left(k\right)$ does not change the sign
difference of $\tau$ nor the difference of upper/lower band numbers.
We thus call $H\left(k\right)$ is stably equivalent to $H^{\prime}\left(k\right)$
if 
\begin{equation}
H\left(k\right)\oplus O\approx H^{\prime}\left(k\right)\oplus O^{\prime},\label{eq:stable}
\end{equation}
denoted as $H\left(k\right)\sim H^{\prime}\left(k\right)$. We regard
the stable equivalent class $\left[H\left(k\right)\right]$ as a topological
phase and achieve the classification by counting out all the $\left[H\left(k\right)\right]$.
Since the homotopic property of $H\left(k\right)$ is determined by
$h_{0}\left(k\right)$ or $h_{0}\left(k\right)\oplus h_{0}^{\prime}\left(k\right)$,
the classification of $\left[H\left(k\right)\right]$ reduces to the
classification of $\left[h_{0}\left(k\right)\right]$ or $\left[h_{0}\left(k\right)\oplus h_{0}^{\prime}\left(k\right)\right]$.
The stable equivalence for subsystems $h_{0}\left(k\right)$ and $h_{0}^{\prime\prime}\left(k\right)$
is reduced from Eq. (\ref{eq:stable}), given by 
\begin{equation}
h_{0}\left(k\right)\oplus o\approx h_{0}^{\prime\prime}\left(k\right)\oplus o^{\prime\prime}.
\end{equation}
We know that $h_{0}\left(k\right)$ is attributed to class A, AI,
or AII, and $o$ is homotopic to the fermionic trivial Hamiltonian
$\sigma$. Therefore, the classification of $\left[h_{0}\left(k\right)\right]$
is directly given by the periodic table of AZ classes Tab. \ref{tab:AZ}.
For independent $h_{0}\left(k\right)$ and $h_{0}^{\prime}\left(k\right)$,
the classification of $\left[h_{0}\left(k\right)\oplus h_{0}^{\prime}\left(k\right)\right]=\left[h_{0}\left(k\right)\right]\oplus\left[h_{0}^{\prime}\left(k\right)\right]$
simply gets doubled.

\section{Characteristic numbers\label{sec:numbers}}

We provide the computation formulas of the Chern number, Chern-Simons
invariant, and Fu-Kane invariant for the single-particle Hamiltonians
$h_{0}\left(k\right)$ and $h_{0}^{\prime}\left(k\right)$\cite{Chiu2016RevModPhys}.
Firstly, the Chern number is defined by 
\begin{equation}
\mathrm{CN}=\int_{\mathrm{BZ}}\det\left(\mathbb{I}+\frac{i}{2\pi}\mathcal{F}\right),
\end{equation}
where $\mathcal{F}=\mathrm{d}\mathcal{A}+\mathcal{A}\land\mathcal{A}$
is the Berry curvature 2-form. Here $\mathcal{A}$ denotes the Berry
connection form of the bands below the gap, and BZ refers to the Brillouin
zone, namely, the $k$-space. It is a topological invariant that measures
the twisting of the energy bands (vector bundle). Secondly, the Chern-Simons
invariant for $d=2m-1$ is a geometrical invariant defined by 
\begin{equation}
\mathrm{CS}=\exp\left(2\pi i\int_{\mathrm{BZ}}\mathrm{cs}_{m}\right).
\end{equation}
The Chern-Simons form reads

\begin{equation}
\mathrm{cs}_{m}=\frac{1}{\left(m-1\right)!}\left(\frac{i}{2\pi}\right)^{m}\int_{0}^{1}\mathrm{d}t\mathrm{Tr}\left(\mathcal{A}\land\mathcal{F}_{t}^{m-1}\right),
\end{equation}
where $\mathcal{F}_{t}=t\mathrm{d}\mathcal{A}+t^{2}\mathcal{A}^{2}$.
With the existence of time-reversal symmetry, $\mathrm{CS}$ takes
discrete values $\pm1$ for $d=3,7$ and then keeps invariant under
the continuous deformation of $\mathcal{A}\left(k\right)$, which
becomes a topological invariant. Thirdly, the Fu-Kane invariant for
$d=2m$ is defined by

\begin{equation}
\mathrm{FK}=\int_{\mathrm{BZ}/2}\frac{1}{m!}\mathrm{Tr}\left(\frac{i}{2\pi}\mathcal{F}\right)^{m}-\int_{\partial\mathrm{BZ}/2}\mathrm{cs}_{m},
\end{equation}
in which $\mathrm{BZ}/2$ refers to a half of Brillouin zone. It also
takes discrete values $\pm1$ for $d=2,6$ with the existence of the
$T$ symmetry and thus becomes a topological invariant.

Finally, we provide the expressions of the Berry connection. Without
loosing generality, we suppose the first $n$ elements of $E=u_{a}^{\dagger}h_{0}u_{a}$
lower than the energy gap. Then the Berry connection form of $h_{0}\left(k\right)$
is given by 
\begin{equation}
\mathcal{A}=\Gamma^{\dagger}u_{a}^{\dagger}\mathrm{d}u_{a}\Gamma,~\Gamma^{\dagger}=\left(\begin{array}{cc}
\mathbb{I}_{n} & 0\end{array}\right).
\end{equation}
Similarly, we suppose that the first $n^{\prime}$ elements of $e^{\prime}=u_{b}^{\dagger}h_{0}^{\prime}u_{b}$
are lower than the energy gap. Then the Berry connection of $h_{0}^{\prime}\left(k\right)$
is given by 
\begin{equation}
\mathcal{A}^{\prime}=\Gamma^{\prime\dagger}u_{b}^{\dagger}\mathrm{d}u_{b}\Gamma^{\prime},~\Gamma^{\prime\dagger}=\left(\begin{array}{cc}
\mathbb{I}_{n^{\prime}} & 0\end{array}\right).
\end{equation}
It is worthy to mention that in previous literature, a bosonic-version
Berry connection is defined via the pseudo-unitary diagonalization
\cite{Shindou2013PRB}, i.e., 
\begin{equation}
\mathcal{A}_{\mathrm{Bose}}=\left(\Gamma^{\dagger},0\right)V^{\dagger}\tau\mathrm{d}V\left(\begin{array}{c}
\Gamma\\
0
\end{array}\right).
\end{equation}
In fact, $\mathcal{A}_{\mathrm{Bose}}$ can be continuously mapped
to $\mathcal{A}$ when $W\left(k\right)$ gradually decreases to zero.
Since continuous mapping does not change the topology, these two Berry
connections lead to the identical topological invariant.


\begin{thebibliography}{65}
\bibitem [{Def()}]{DefRe}%
	\BibitemOpen
	\href@noop {} {}\bibinfo {note} {A deformation retraction
		$F_{\epsilon}:G\rightarrow G$ satisfies that (1) $F_{0}:G\rightarrow G$ is an
		identity mapping; (2) $F_{1}:G\rightarrow G_{1}$ is a retraction mapping; (3)
		$F_{\epsilon}:G_{1}\rightarrow G_{1}$ is identity mapping for each
		$\epsilon$.}\BibitemShut {Stop}%
\end{thebibliography}
\end{document}